# Evaluation of Machine Learning Reconstruction Techniques for Accelerated Brain MRI Scans

Jonathan I. Mandel [1], Shivaprakash Hiremath [2], Hedyeh Keshtgar [3], Timothy Scholl [4], Sadegh Raeisi [3]

[1] Humber River Hospital, Toronto, Canada
[2] Department of Medical Imaging, Toronto Western Hospital, University of Toronto, Toronto, Canada
[3] Foqus Technologies Inc., Kitchener, Canada
[4] Department of Medical Biophysics, Robarts Research Institute, Western University, London, Canada

**Abstract**

This retrospective–prospective study evaluated whether a deep learning-based MRI reconstruction algorithm (DeepFoqus-Accelerate; Foqus, 510(k) K241982) can preserve diagnostic quality in brain MRI scans accelerated up to fourfold, using both public and prospective clinical data. The study included 18 healthy volunteers (scans acquired at 3T, January 2024–March 2025), as well as selected fastMRI public datasets with diverse pathologies. Phase-encoding-undersampled 2D/3D T1, T2, and FLAIR sequences were reconstructed with DeepFoqus-Accelerate and compared with standard-of-care (SOC). Three board-certified neuroradiologists and two MRI technologists independently reviewed 36 paired SOC/AI reconstructions from both datasets using a 5-point Likert scale, while quantitative similarity was assessed for 408 scans and 1224 datasets using Structural Similarity Index (SSIM), Peak Signal-to-Noise Ratio (PSNR), and Haar wavelet-based Perceptual Similarity Index (HaarPSI). No AI-reconstructed scan scored below 3 (minimally acceptable), and 95% scored ≥4. Mean SSIM was 0.95±0.03 (90% cases >0.90), PSNR >41.0 dB, and HaarPSI >0.94. Inter-rater agreement was slight to moderate. Rare artifacts did not affect diagnostic interpretation. These findings demonstrate that DeepFoqus-Accelerate enables robust fourfold brain MRI acceleration with 75% reduced scan time, while preserving diagnostic image quality and supporting improved workflow efficiency.[1]

---

**Introduction**

MRI is a cornerstone of diagnostic imaging, providing exceptional soft tissue contrast and enabling detailed characterization of neurological, vascular, and musculoskeletal diseases. Despite its clinical utility, prolonged acquisition times remain a primary limitation of MRI. Extended scan durations increase patient discomfort, particularly in vulnerable populations such as children, claustrophobic patients, and those unable to remain still, all of which elevate the risk of motion artifacts that can degrade image quality and necessitate repeat imaging[1,2]. Furthermore, longer scan times reduce throughput capacity, contribute to scheduling bottlenecks, and increase operational costs, impacting healthcare delivery efficiency.

To address these challenges, accelerated MRI acquisition techniques are under development. Traditional methods, including parallel imaging and compressed sensing, while clinically implemented, typically achieve acceleration factors up to two and may compromise signal-to-noise ratio or introduce artifacts at higher accelerations, limiting their effectiveness in routine practice[3-6].

Recent advances in machine learning, especially deep learning (DL), promise to overcome these limitations by enabling high-quality MRI reconstruction from heavily undersampled data[7]. DL-based MRI reconstruction methods can be data-driven, learning mappings from undersampled or aliased inputs to high-fidelity outputs using large datasets, or physics-informed, incorporating MR signal encoding to improve robustness. The main data-driven categories are image-domain, k-space-domain, and hybrid reconstructions, each offering distinct advantages. Image-domain methods learn direct pixel-wise mappings from aliased to high-quality images, while k-space methods preserve high-frequency details[8].

DeepFoqus-Accelerate is an FDA-cleared, k-space-based DL MRI reconstruction algorithm that leverages proprietary neural networks trained on extensive MRI datasets to enable robust fourfold acceleration of brain MRI while preserving diagnostic quality. Integrating AI tools into routine clinical practice requires rigorous validation to confirm performance equivalency to standard-of-care (SOC) imaging, encompassing both expert review and quantitative metrics[9].

This study aims to evaluate DeepFoqus-Accelerate's performance on a mixed retrospective-prospective dataset combining publicly available fastMRI data with pathological cases and prospectively acquired scans from healthy volunteers. We hypothesize that

DeepFoqus-Accelerate preserves diagnostic integrity of accelerated brain MRI acquisition, supporting improved workflow efficiency and patient experience.

**Materials and Methods**

*Study Design and Population*

A mixed retrospective-prospective design was employed. Retrospective data were sourced from the publicly available fastMRI dataset[10], encompassing multi-coil raw k-space brain MRI data acquired across 1.5T and 3T Siemens platforms, representing a wide range of ages, sexes, and clinical pathologies. Prospective data collection included 18 healthy adult volunteers (8 men, 10 women; mean age 35.5 years, range 22–66) recruited between January 2024 and March 2025 after institutional review board approval. Scans were de-identified by excluding all personal identifiers from the DICOM headers, retaining only age and sex. Written informed consent was obtained from all participants. The prospective cohort size was chosen for feasibility and to ensure evaluation datasets diversity, meeting the minimum required to reliably assess the software's performance.

*Imaging Protocols*

The fastMRI dataset included fully-sampled 2D axial T1, T2, and FLAIR sequences from 1.5T and 3T Siemens scanners[10]. Prospectively, participants underwent standardized SOC and fourfold accelerated brain MRI scanning on a 3T GE Discovery MR750 system. Acquisitions included a broad array of sequences: 2D axial, coronal, and sagittal T1 spin echo (SE), T2 fast spin echo (FSE), and T2 FLAIR sequences, along with 3D axial T1 BRAVO, sagittal T2, and coronal FLAIR sequences. Acquisition parameters are detailed in Supplementary Table 1. For this study, accelerations were simulated by retrospectively reducing phase-encoding steps to achieve 2x, 3x, and 4x undersampling ( ~50%, 66%, and 75% scan time reductions). Thirty-six paired datasets were selected for qualitative review, using random selection within categories of scan features to ensure diversity and provide a representative evaluation.

*Image Reconstruction*

DeepFoqus-Accelerate version 1.1 (Foqus; 510 K clearance K241982) was utilized to reconstruct undersampled k-space data. This FDA-cleared algorithm uses a proprietary deep neural network architecture trained on a large, heterogeneous MRI dataset external to the study population, minimizing bias.

*Qualitative Assessment*

Qualitative image assessments were performed by five experienced raters: three board-certified neuroradiologists (with 10–15 years of experience) and two MRI technologists (with 10 and 18 years of experience), all independent of the software development team. Thirty-six anonymized paired datasets (SOC and AI-reconstructed accelerated scans) were reviewed. Overall image quality was scored on a 5-point Likert scale (1 = non-diagnostic, 5 = identical to SOC), focusing on diagnostic utility and artifact presence (Supplementary Table 2). Scores and reviewer notes were independently collected without consensus reading. Inter-rater reliability was calculated using weighted Cohen's kappa, and consistency of scoring patterns was analyzed via Spearman's correlation coefficient with a significance threshold of $p < 0.05$.

*Quantitative Assessment*

Four hundred and eight scans with multiple acceleration rates yielded 1224 sets evaluated using quantitative metrics, including Structural Similarity Index Measure (SSIM), Peak Signal-to-Noise Ratio (PSNR), and Haar wavelet-based Perceptual Similarity Index (HaarPSI). These objective image similarity metrics are strongly correlated with radiologist-perceived quality[11,12]. Confidence intervals were calculated, and evaluation was performed across sequences (T1, T2, and FLAIR).

## Results

*Qualitative Evaluations*

Across all expert readers, no AI-reconstructed images were rated below a diagnostic quality threshold (all scores ≥3), with 95% scoring ≥4, denoting high diagnostic confidence (Figure 1). Mean and median scores were 4.38±0.35 and 4.4, respectively, reflecting consistent image quality perception. Minor inter-reader scoring variability was observed: weighted Cohen's kappa values ranged from slight to moderate agreement (κ = -0.18 to 0.46) with fair overall concordance (Figure 2b). Spearman correlation (Figure 2a) highlighted the strongest concordance between reviewers 1, 2, and 4 ($p < 0.01$); reviewer 2 applied stricter criteria, though differences did not impact diagnostic interpretations. Representative images illustrate high-fidelity reconstructions; artifacts such as wrap-around and motion-related distortions, originally present in the reference images, did not impede lesion detection or anatomical delineation in the reconstructed images, although they may have contributed to lower quantitative metrics' scores. (Figure 4)

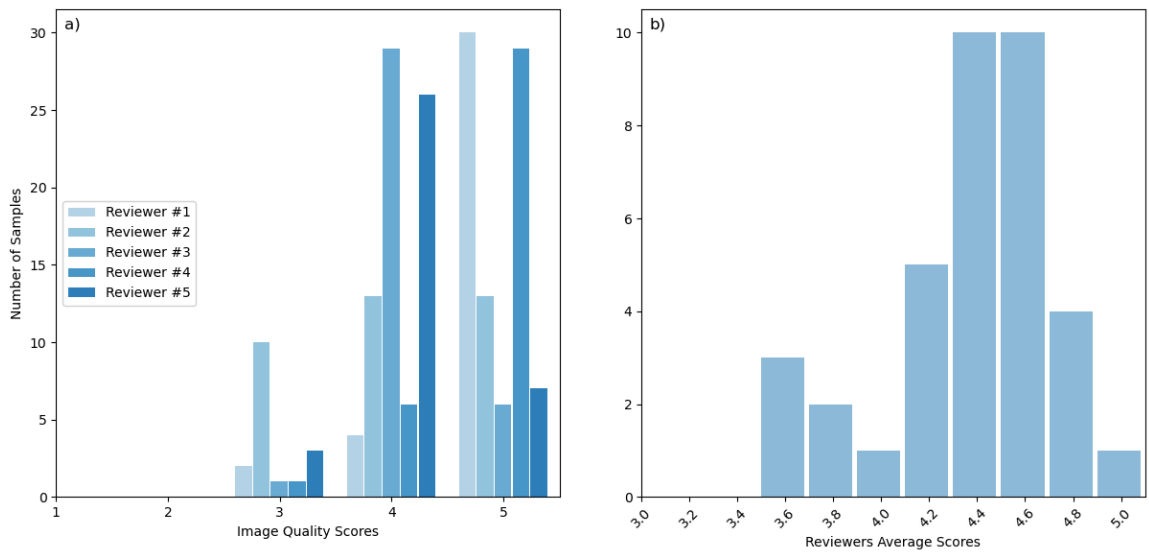

**Figure 1:** Histograms representing **(a)** the distribution of reviewer scores and **(b)** the average scores for 36 image sets, with AI-reconstructed brain MRI scans rated against standard of care (SOC) images using a 5-point Likert scale.

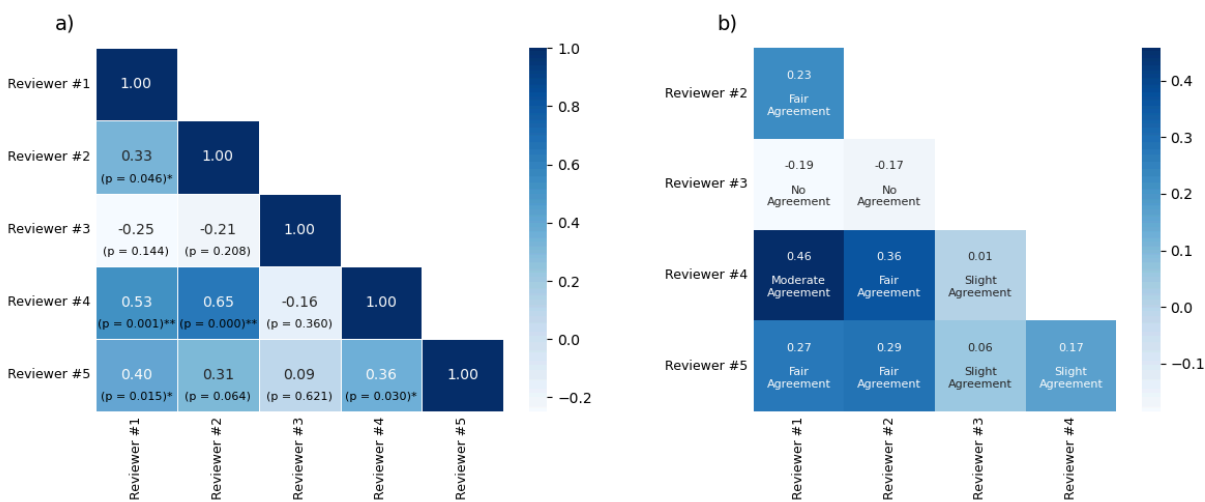

**Figure 2: (a)** Pair-wise Spearman's correlation matrix of reviewer scores, with corresponding p-values. Statistical significance is indicated by asterisks: ** for $p \leq 0.01$ and * for $p \leq 0.05$. **(b)** The $\mathcal{K}$ scores for all pairs of reviewers.

*Quantitative Evaluations*

Quantitative evaluation showed high structural similarity between DeepFoqus-Accelerate reconstructions and SOC images across all sequences (Table 1). For the full quantitative dataset, the mean SSIM was $0.959 \pm 0.034$, and within the qualitative subset it was $0.952 \pm 0.036$. PSNR averaged above 41 dB, and HaarPSI values exceeded 0.94. Over 90% of AI reconstructions demonstrated SSIM >0.90, indicating consistently high similarity to the reference images. A small number of outliers showed reduced scores, aligned with reader-noted artifacts, prompting further review of specific acquisition or reconstruction variables.

**Table 1: Quantitative Assessment**

| Evaluation Datasets | SSIM | PSNR (dB) | HaarPSI |
|---|---|---|---|
| Quantitative Dataset | 0.959 ± 0.034 (95% CI: 0.957, 0.960) | 41.738 ± 4.58 (95% CI: 41.482, 41.995) | 0.954 ± 0.030 (95% CI: 0.953, 0.956) |
| • T1 | 0.963 ± 0.025 (95% CI:0.961, 0.965) | 42.201 ± 3.99 (95% CI:41.853, 42.549) | 0.960 ± 0.023 (95% CI:0.958, 0.962) |
| • T2 | 0.968 ± 0.024 (95% CI:0.965, 0.971) | 42.427 ± 4.87 (95% CI:41.904, 42.950) | 0.965 ± 0.019 (95% CI:0.963, 0.967) |
| • FLAIR | 0.944 ± 0.045 (95% CI:0.939, 0.949) | 40.502 ± 4.80 (95% CI:40.017, 40.988) | 0.938 ± 0.041 (95% CI:0.933, 0.942) |
| Qualitative Dataset | 0.952 ± 0.036 (95% CI: 0.940, 0.964) | 41.159 ± 3.67 (95% CI: 39.916.310,42.401) | 0.944 ± 0.031 (95% CI: 0.933, 0.954) |

Summary of mean ± standard deviation with 95% Confidence Interval (CI) for the Structural Similarity Index (SSIM), Peak Signal-to-Noise Ratio (PSNR), and Haar wavelet-based Perceptual Similarity Index (HaarPSI) scores.

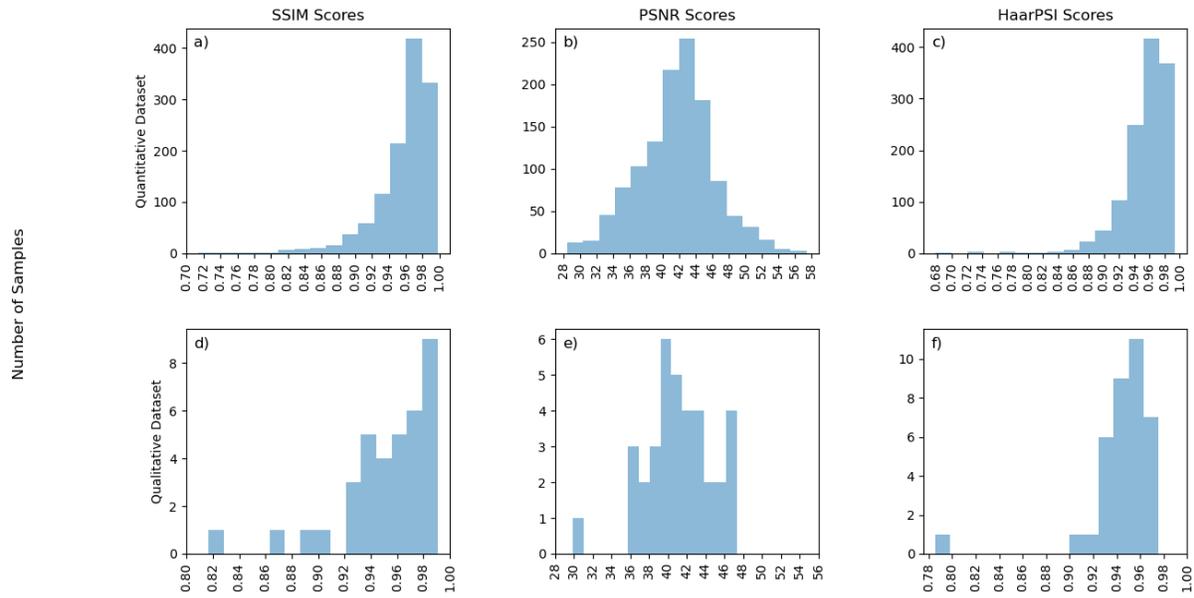

**Figure 3:** Distribution of Structural Similarity Index (SSIM), Peak Signal-to-Noise Ratio (PSNR), and Haar wavelet-based Perceptual Similarity Index (HaarPSI) scores for DeepFoqus-Accelerate reconstructions: **(a–c)** show results across 408 samples at 2x, 3x, and 4x acceleration, and **(d–f)** present distributions for the 36 image sets evaluated by reviewers.

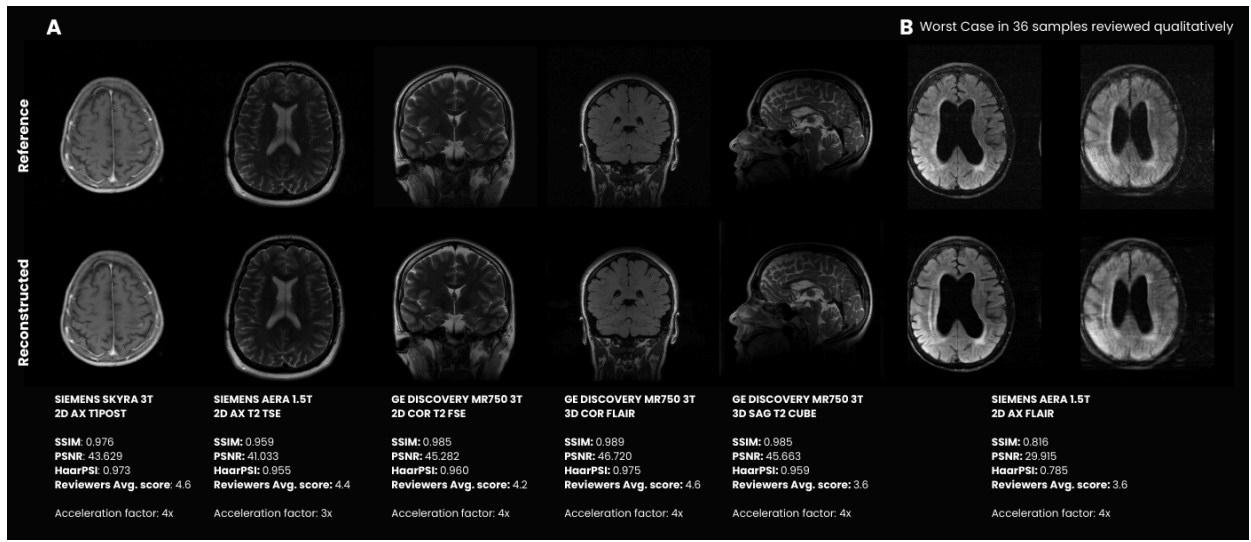

**Figure 4: (A–B)** Representative standard-of-care (SOC) images (first row) and DeepFoqus-Accelerate reconstructions from accelerated scans (second row), with corresponding quantitative and qualitative scores presented in the third row. Panel **(B)** shows two slices of the worst-case scenario in the qualitative dataset, characterized by wrap-around and motion artifacts.

**Discussion**

This evaluation of DeepFoqus-Accelerate demonstrates that this FDA-cleared k-space–based DL reconstruction software can reliably enable up to fourfold accelerated brain MRI acquisition without compromising diagnostic image quality. Both expert review and quantitative image similarity metrics confirm that AI-reconstructed images are clinically equivalent to fully sampled standards.

Our results align with emerging literature showing that DL-accelerated MRI robustly matches conventional imaging without sacrificing diagnostic confidence even in demanding applications such as stroke imaging, pediatric assessments, and combined 2D/3D protocols[13-15].

The strength of DeepFoqus-Accelerate lies in its proprietary network design and k-space–based reconstruction, which together enable MRI acquisition up to fourfold faster while preserving image details, maintaining anatomical fidelity, and supporting reliable diagnostic interpretation across a range of clinical scenarios. While infrequent, minor intensity nonuniformity or occasional artifacts, consistent with known limitations of DL–based reconstructions[16-18], did not impede diagnostic interpretation and highlighted areas needing future algorithm refinement.

A unique strength of this study was the inclusion of a hybrid dataset, encompassing publicly available data and prospective real-world data with a multi-expert qualitative review and rigorous quantitative analysis. The slight-to-moderate inter-reader agreement we observed is consistent with known subjectivity in image interpretation, reinforcing the importance of multi-reader panels to validate clinical utility. The dual use of expert Likert ratings and robust quantitative metrics (SSIM, PSNR, HaarPSI) provides complementary, convergent validation of image quality, strengthening the rigour of these findings. This multi-pronged methodology exceeds the scope of many prior, single-center investigations, conferring greater generalizability compared to earlier studies.

Clinically, the scan-time reduction of up to 75% translates into tangible benefits, including less patient discomfort, decreased risk of motion artifacts, fewer sedation requirements, improved access, and greater throughput. Furthermore, our results support the suitability of these AI-reconstructed images for advanced quantitative tasks such as automated volumetric analysis and lesion segmentation, reinforcing the broad clinical utility of the technique with no compromise in critical diagnostic features.

Despite rigorous assessments, some inter-reader variability was noted, likely reflecting inherent reader subjectivity and subtle differences that may be amplified with advanced acceleration and should motivate ongoing refinement of reconstruction algorithms and reader training practices when deploying new AI tools.

The limitations of our study include a relatively small, single-center prospective cohort and the absence of prospective multi-vendor scanner diversity, which are important next steps. As AI-accelerated MRI sees wider adoption, future studies should embrace multicenter and multi-vendor designs, and address an even wider range of clinical indications to further validate and expand the use of AI-accelerated MRI in routine practice.

**Conclusion**

The DeepFoqus-Accelerate FDA-cleared DL algorithm reliably supports fourfold accelerated brain MRI acquisition while maintaining diagnostic quality equivalent to established standards. Its application promises significant improvements in patient throughput, comfort, and MRI suite efficiency without compromising diagnostic performance, positioning it as a valuable tool for advancing neuroimaging clinical practice.

**Acknowledgment**

This study was supported in part by the Ontario Centre of Innovation (OCI). The authors thank the MRI technologists at Robarts Research Institute for their help with volunteer recruitment and data collection, Dr. Trevor Wade for valuable input, Dr. Barbara Murray for ethics documentation, and Dr. Sean A. Kennedy, Trevor Szekeres, and Anthony D. Zagar for their contributions.

**Supplemental Material:**

**Undersampling Pattern:**

The undersampling pattern (Figure 1) maintains the specified center fraction and undersamples the remaining lines to obtain the defined acceleration rate, such that lines are preserved in an equispaced manner (i.e., each preserved phase encoding line is separated by an equal number of omitted phase encoding lines).

To apply 4X acceleration for example, the target undersampling rate is 4, which means ¾ of the total number of acquisition lines (i.e., maximum number of lines if the acquisition were fully sampled) in the undersampling dimension need to be removed. In other words, ¼ of phase encoding lines are preserved.

In addition, the fraction of the center of the k-space which is maintained is 8%. This means that all the central lines of the k-space up to 8% are preserved. For example, if there are a total of 100 lines in the k-space, a band of the 8 fully-sampled lines is preserved at the center and undersampling is applied outside this band. The center fraction ensures that all of the low frequency data at the center of the k-space is preserved.

Figure 1: Visual representation of accelerated k-space acquisition and the corresponding mask
The white lines (binary mask value of one), correspond to the phase encoding lines that are preserved, and the black lines (binary mask value of zero) correspond to the phase encoding lines that are not collected.

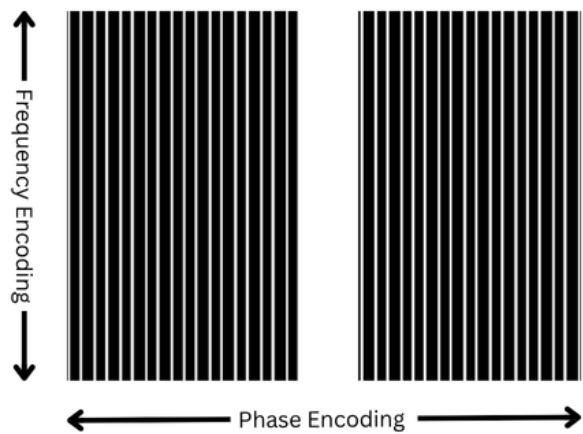

**Image Acquisition Parameters and Scan Times:**

The parameters in Table 1 were used for scans performed on the GE Discovery MR750 3T MRI scanner. All scans were conducted with the number of averages (NEX) and phase FOV set to 1.

**Table 1:** Acquisition Parameters of GE Scans

| Sequence | TR/TE/TI | Flip angle | FOV (mm) | Slice thickness (mm) | Acquisition matrix | Freq DIR | Scan Time (SOC) | Scan Time (Accelerated) |
|---|---|---|---|---|---|---|---|---|
| 2D AX T1 SE | 600/Min Full 667/Min Full | 90 | 210 | 4 | 280 × 224 | A/P | 2.0 min and 22.85 sec | 0.0 min and 41.38 sec |
| 2D AX T2 FSE | 6198/110 | 142 111 | 210 | 4 | 300 × 300 | A/P | 1.0 min and 45.57 sec | 0.0 min and 31.19 sec |
| 2D AX T2 FLAIR | 10000/120/2571 10000/119.9/2571 | 160 | 210 | 4 | 300 × 240 | A/P | 4.0 min and 10.4 sec | 1.0 min and 30.4 sec |
| 2D Coronal T1 SE | 700/Min Full 767/Min Full | 90 | 210 | 4 | 280 × 224 | S/I | 2.0 min and 46.65 sec | 0.0 min and 46.55 sec |
| 2D Coronal T2 FSE | 7214/110 | 142 111 | 210 | 4 | 300 × 300 | S/I | 2.0 min and 2.84 sec | 0.0 min and 36.27 sec |
| 2D Coronal T2 Flair | 8000/120/2354 8000/119.9/2354 | 160 | 210 | 4 | 300 × 240 | S/I | 4.0 min and 56.6 sec | 1.0 min and 44.56 sec |
| 2D Sagittal T1 SE | 600/Min Full 650/Min Full | 90 | 210 | 4 | 280 × 224 | S/I | 2.0 min and 22.84 sec | 0.0 min and 40.35 sec |
| 2D Sagittal T2 FSE | 6196/110 | 142 111 | 210 | 4 | 300 × 300 | S/I | 1.0 min and 45.53 sec | 0.0 min and 31.18 sec |
| 2D Sagittal T2 Flair | 10000/120/2571 10000/119.9/2571 | 160 | 210 | 4 | 300 × 240 | S/I | 4.0 min and 10.4 sec | 1.0 min and 30.4 sec |
| 3D AX T1 BRAVO | 450 | 10 | 245 | 1 | 256 × 256 | A/P | 8.0 min and 15.58 sec | 2.0 min and 19.52 sec |
| 3D Sagittal T2 | 2500/Maximum | 90 | 256 | 1 | 256 × 256 | S/I | 14.0 min and 49.02 sec | 3.0 min and 15.90 sec |

| 3D Coronal FLAIR | 5002/103.187/1521 | 90 | 256 | 1 | 256 ╳ 256 | S/I | 18.0 min and 28.53 sec | 5.0 min and 38.21 sec |

**Qualitative Assessment:**

Reviewers scored the overall image quality of the AI-reconstructed accelerated scan compared to the SOC from a diagnostic point of view using the predefined 1–5 scale in Table 2.

**Table 2:** Image Quality Scoring Scale for Reconstructed Brain MRI

| Scale | Description |
|---|---|
| 1 | Score "1" means that critical features are completely missing or an extra artifact is added to the reconstructed image. |
| 2 | Score "2" means that the reconstructed image is severely distorted by additional artifacts or noise, such as motion. It affects the radiologist's ability to interpret the information presented in the image. |
| 3 | Score "3" means that the image meets the minimum acceptable quality standards. The reconstructed image may contain minor artifacts or noise, but they do not affect the ability of the radiologist to interpret the information presented in the image. |
| 4 | Score "4" means that despite some minor differences between the two images, the reconstructed image is an almost perfect reconstruction of the original. Important features are preserved with no significant information or artifacts added to or removed from the image. |
| 5 | Score "5" means that the reconstructed image is exactly the same as the original image from the diagnostic point of view. All important features are preserved with no information or artifacts added to the image. |